\documentclass[a4paper]{ESASPCS13Style}
\usepackage{epsfig}
\usepackage{color}

\definecolor{red}{rgb}{1.00,0.00,0.00}
\definecolor{blue}{rgb}{0.00,0.00,1.00}
\definecolor{firebrick}{rgb}{0.70,0.13,0.13}
\definecolor{darkorange}{rgb}{1.00,0.55,0.00}
\definecolor{purple}{rgb}{0.63,0.13,0.94}
\definecolor{yellow4}{rgb}{0.55,0.55,0.00}
\definecolor{midnightblue}{rgb}{0.10,0.10,0.44}

\newcommand{\imu}{\ensuremath{m}}

\newcommand{\nmu}{\ensuremath{M}}

\newcommand{\Ntot}{\ensuremath{N}}

\newcommand{\Imu}{\ensuremath{I_{\imu}}}

\newcommand{\HImu}{\ensuremath{\FT{I}_{\imu}}}

\newcommand{\HCImu}{\ensuremath{\conj{\FT{I}}_{\imu}}}

\newcommand{\mueffmu}{\ensuremath{\tilde{\mu}_{\imu}}}

\newcommand{\var}[1]{{\ensuremath{\sigma^2_{#1}}}}
\newcommand{\sig}[1]{{\ensuremath{\sigma_{#1}}}}

\newcommand{\Ap}{\ensuremath{A}}
\newcommand{\lp}{\ensuremath{l}}

\newcommand{\fobs}{\ensuremath{f}}
\newcommand{\Rstar}{\ensuremath{R}}
\newcommand{\msum}[1]{\ensuremath{\sum_{#1=1}^{\nmu}}}

\newcommand{\wmu}{\ensuremath{w_\imu}}

\newcommand{\tmean}[1]{\ensuremath{\left\langle #1\right\rangle}}

\newcommand{\Hfobs}{\ensuremath{\FT{\fobs}}}
\newcommand{\HCfobs}{\ensuremath{\conj{\FT{\fobs}}}}
\newcommand{\FT}[1]{\ensuremath{\hat{#1}}}
\newcommand{\conj}[1]{\ensuremath{#1^\ast}}

\newcommand{\xph}{\ensuremath{x_\mathrm{ph}}}

\newcommand{\COBOLD}{{\sc CO$^5$BOLD}}
\newcommand{\pun}[1]{\mbox{\rm\,#1}}

\newcommand{\Teff}{\ensuremath{T_{\mathrm{eff}}}}
\newcommand{\logg}{\ensuremath{\log g}}

\newcommand{\beq}{\begin{equation}}
\newcommand{\eeq}{\end{equation}}

\newcommand{\nobm}{\ensuremath{\mathrm{N}_\mathrm{obm}}}
\newcommand{\dirms}{\ensuremath{\delta I_\mathrm{rms}}}
\newcommand{\tauc}{\ensuremath{\tau_\mathrm{c}}}
\newcommand{\Hpsurf}{\ensuremath{H_\mathrm{p}^\mathrm{surf}}}
\newcommand{\eref}[1]{\mbox{(\ref{#1})}}

\begin{document}

\title{Hydrodynamical simulations of convection-related stellar micro-variability}

\author{Fredrik Svensson \and\
  Hans-G{\"u}nter~Ludwig}
\institute{Lund Observatory, Box~43, 22100~Lund, Sweden}

\maketitle 

\begin{abstract}
 We used a series of \COBOLD\
 hydrodynamical model atmospheres covering stellar objects from white
 dwarfs to red giants to derive theoretical estimates of the
 \textit{photometric\/} and \textit{photocentric stellar
 variability\/} in wavelength-integrated light across the
 Hertzsprung-Russell diagram. We validated our models against solar
 measurements from the {\sc Soho/Virgo} instrument. Within our set of
 models we find a systematic increase of the photometric as well as
 photocentric variability --- which turn out to be closely connected
 --- with decreasing surface gravity. The estimated absolute levels of
 the photocentric variability do not affect astrometric observations
 on a precision level expected to be achieved by the GAIA mission ---
 with the exception of close-by giants. The case of supergiants
 remains to be investigated. In view of the ongoing debate about the
 photometric non-detection of p-modes in Procyon by the Canadian {\sc
 Most} satellite we remark that we obtain a factor of $\approx 3$ in
 amplitude between the granular background ``noise'' in the Sun and
 Procyon. This statement refers to a particular representation of
 temporal power spectra as discussed in Sect.~\ref{s:trendphoto}.
\end{abstract}

\section{Introduction}

\begin{figure}[t!]
\begin{center}
\includegraphics[width=0.8\hsize]{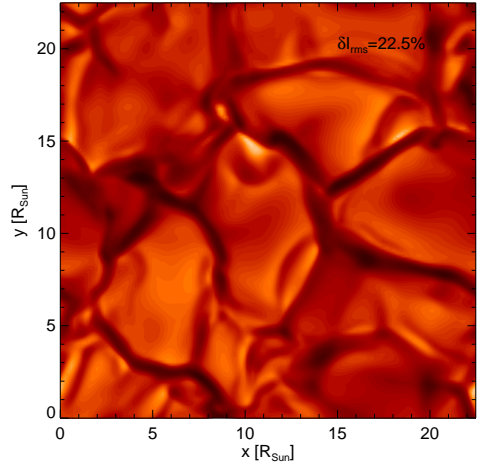}
\caption{Typical granular intensity pattern from the red giant model
sequence. Note the spatial scales. 
\label{f:intensity}}
\end{center}
\end{figure}

\begin{figure}[t!]
\begin{center}
\includegraphics[width=0.8\hsize]{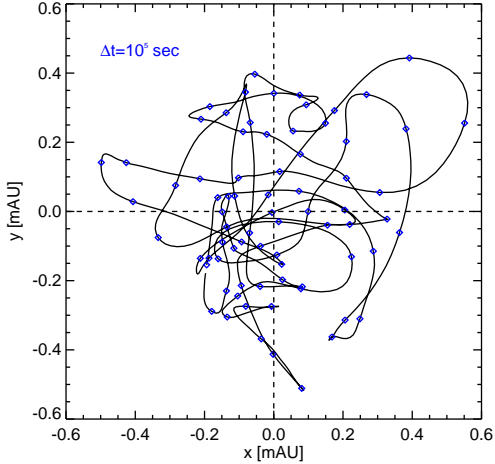}
\caption{A statistical realisation of the motion of the photocenter in the plane
of the sky based on data from the red giant simulation. The time between two
diamonds amounts to $10^5\pun{s}$.
\label{f:xphpath}}
\end{center}
\end{figure}

The presence of a small-scale time-dependent granulation pattern on
the surfaces of late-type stars (see Fig.~\ref{f:intensity}) leads to
low-level temporal fluctuations in a star's total luminosity, apparent
position given by its photocenter, (see Fig.~\ref{f:xphpath}), and
spectroscopically measured radial velocity. During recent years this
kind of micro-variability got into the focus of research since it
forms an important noise source in searches for solar-like
oscillations and exoplanets. This is particularly the case for already
operational or upcoming satellite mission like {\sc Most}, {\sc
Corot}, and {\sc Kepler}; granulation induced micro-variability might
also become a limiting factor for precision astrometry missions like
{\sc SIM} and {\sc GAIA} (\cite{Hatzes02}, \cite{Green+al03},
\cite{Aigrain+al04}, \cite{Matthews+al04}).  We performed
radiation-hydrodynamics simulations for stellar atmospheres to derive
\textit{theoretical estimates\/} of the convection-related photometric
and photocentric variability across the Hertz\-sprung-Russell
diagram. To our knowledge the work of \cite*{Trampedach+al98} is the
only previous example of such a theoretical effort who studied the
brightness and radial velocity variability in the Sun, $\alpha$~Cen~A,
and Procyon~A.

\section{\COBOLD\ rad.-hydrodynamics simulations}

We used the radiation-hydrodynamics code \COBOLD\ (for further
information about the code and applications see \cite{fre02}, and
\cite{WFSLH04}) to construct a series of 3D Cartesian ``local-box''
model atmospheres.  Table~\ref{t:models} summarises the model
properties. We calculated three solar models which differ in numerical
details. They are used as basic reference and for validating our
models by comparison with {\sc Soho/Virgo} observational data. The two
models ``Procyon~A'' and ``$\xi$~Hydrae'' have parameters close to the
actual parameters of the stars they are named after. The other models
are not intended to represent particular stars but are generic
spanning a large range primarily in surface gravity. Two models are intended to
investigate effects of metallicity. Note, that the variation in
effective temperature among the models is not very large, i.e. the set
of models does not fully cover the parameter space. While not apparent
from the table significant effort was invested to follow the evolution
of the models over long periods of time to ensure statistically
representative results.

\begin{table*}
\caption{\COBOLD\ radiation-hydrodynamics model atmospheres: ``Model'' is a
  model's nickname used in this paper, \Teff\ the effective temperature,
  \logg\ the gravitational acceleration, \Rstar\ an assumed stellar radius
  (not intrinsic to the simulation proper), \lp\ the linear horizontal size of
  the computational box, \tauc\ the sound crossing time over \Hpsurf,
  \Hpsurf\ the pressure scale height at Rosseland optical depth unity, \dirms\
  the relative spatial white light intensity contrast, \nobm\ the number of
  equivalent frequency points considered in the solution of the radiative
  transfer equation (RTE), \nmu\ the number of inclinations used in the
  discretisation of the RTE, ``Modelcode'' an internal identifier of the model
  sequence.
\label{t:models}}
\begin{tabular}[t]{|l|l|l|l|l|l|l|l|l|l|l|l|}
\hline
Model &  $T_\mathrm{eff}$ & \logg  & Radius \Rstar & Box size~\lp 
& $\tau_\mathrm{c}$ & $H_{\mathrm{p}}^{\mathrm{surf}}$ & \dirms & \nobm & \nmu & Modelcode\\
& [K] & [$\mathrm{cm/s^2}$] & [Mm] & [Mm] & [s] & [Mm] &        &       &      &       \\
\hline
{\color{darkorange}Sun1}      & 5\,780  & 4.44 & 696     & 5.6     & 17.8 & 0.141 & 0.172 & 5 & 3  & d3gt57g44n53\\
{\color{red}Sun2}             & 5\,740  & 4.44 & 696     & 18.0    & 17.8 & 0.141 & 0.156 & 1 & 2 & d3gt57g44sg6\\
{\color{blue}Sun3}            & 5\,760  & 4.44 & 696     & 11.2    & 17.8 & 0.141 & 0.176 & 5 & 3 & gt57g44n67\\
{\color{blue}White Dwarf}     & 12\,000 & 8.00 & 8.9     &
7.5 [km] & 10.6 [ms] & 133 [m]& 0.173 & 1 & 2 & d3t120g80wd1\\
{\color{midnightblue}$[\mathrm{M}/\mathrm{H}]=0.0$} & 5\,170 & 4.44 &
696 & 4.85 & 16.9 & 0.128 & 0.108 & 5 & 2 & d3gt50g44n01\\
$[\mathrm{M}/\mathrm{H}]=-2.0$ & 4\,730 & 4.44 & 696     & 4.85    & 17.2 & 0.118 & 0.047 & 6 & 3 & d3ot50g44n01\\
{\color{purple}Procyon A}     & 6\,540  & 4.00 & 1\,460  & 29.54   & 49.6 & 0.391 & 0.212 & 1 & 2 & d3gt65g40n2\\
{\color{yellow4}$\xi$ Hydrae} & 4\,880  & 2.94 & 7\,340  & 147.5   & 513  & 3.74  & 0.180 & 1 & 2 & d3gt50g29n01\\
{\color{red}Cepheid}          & 4\,560  & 2.00 & 21\,000 & 2\,125  & 4\,310 & 30.2 & 0.185 & 1 & 2 & d3t50g20mm00n2\\
{\color{firebrick}Red Giant}  & 3\,680  & 1.00 & 66\,300 & 15\,750 & 40\,700 & 242 & 0.220 & 1 & 2 & d3t38g10mm00n2\\
\hline
\end{tabular}
\end{table*}

\section{From simulation boxes to stars}

The \COBOLD\ simulations provide a statistically representative,
rectangular patch (or ``tile'') of the emergent radiation field and
its temporal evolution. In order to derive disk-integrated, observable
quantities we have to extrapolate this information to the whole
visible stellar hemisphere. To this end we envision the stellar
surface being tiled by a possibly large number of simulation
patches. Being interested in integrated quantities we can ignore the
spatial structure within in a patch and only need to consider averages
of the emergent intensity~\Imu\ over the patch surface as a function
of time and inclination $\mueffmu=\cos(\vartheta_\imu)$. Due to limitations
of computing resources the discretisation of the radiation field in
solid angle is rather coarse, and usually we have only a total number
of inclinations~\nmu\ (see Tab.~\ref{t:models}) of two or three
available.

\begin{figure}[t!]
\begin{center}
\includegraphics[width=0.6\hsize,angle=90]{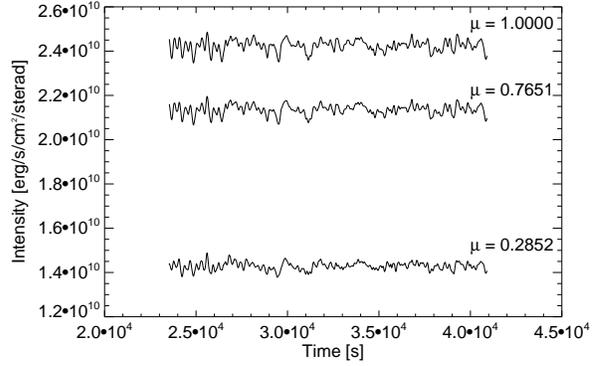}
\caption{Intensity time series for the three inclinations of model
Sun1 showing residual fluctuations and overall limb-darkening.\label{f:timeseries}}
\end{center}
\end{figure}

Figure~\ref{f:timeseries} shows an example of a resulting time series
with $\nmu=3$. Clearly visible is the effect of limb-darkening and
residual intensity fluctuations after averaging over the patch's
surface. The crucial assumption now is that the size of a patch is
large enough so that its emission can be considered as statistically
independent of all other patches tiling the surface. Moreover, we
assume that all patches share the same statistics given by the
statistics of the simulated patch. With these assumptions \cite*{L04}
derived expressions for the statistics of disk-integrated observables
which we summarise below.

\paragraph{Statistics of the photometric variability.} 
We characterise the photometric variability by the temporal power
spectrum of the relative fluctuations of the observable flux~\fobs.
The expectation value of a frequency component denoted by
$\Hfobs(\nu)$ is then given by
\beq
\frac{\tmean{\Hfobs\HCfobs}}{\tmean{\fobs}^2}
=\Ntot^{-1}\frac{\msum{\imu}\wmu\mueffmu^2\tmean{\HImu\HCImu}}%
{\left(\msum{\imu}\wmu\mueffmu\tmean{\Imu}\right)^2}.
\label{e:statisticsf}
\eeq 
Angular brackets denote expectation values, an asterisk the conjugate
complex. The sums are discrete analogs of integrals over one half of
the total solid angle where \wmu\ is the integration weight. The
numerator expresses that a frequency component of the
resulting power spectrum is a $\mueffmu^2$ weighted integral of
the frequency components of the individual intensity power spectra. The
denominator is essentially the square of the average observable
flux. The spectral power density scales inversely proportional to the
number of patches~\Ntot\ tiling the visible hemisphere which is
given by
\beq
\Ntot\Ap=2\pi\Rstar^2,
\eeq
where \Ap\ is the patch surface area and \Rstar\ is the stellar
radius. \Rstar\ is assumed for the star the atmosphere model is
associated with.  In this paper we used plausible but to some extend
arbitrary values for the stellar radii. For possible later adjustments
of the radius one should keep in mind that the power of the
photometric fluctuations scales as $\Rstar^{-2}$. The power spectra
presented in this paper are given as spectral density normalised so
that the integral between zero and the Nyquist frequency corresponds
to the variance of the original signal.

\paragraph{Statistics of the photocentric displacement.} We characterise
the absolute photocentric displacement along an arbitrary axis~\xph\ by
its standard deviation~\sig{\xph}\ given to good approximation by
\begin{eqnarray}
\sig{\xph}&\approx&
\frac{\Rstar}{\sqrt{6}}\frac{\sig{\fobs}}{\tmean{\fobs}}
=\frac{\lp}{\sqrt{12\pi}}\,\Ntot^\frac{1}{2}\frac{\sig{\fobs}}{\tmean{\fobs}}\nonumber\\
&=&\frac{\lp}{\sqrt{12\pi}}\,
\frac{\left(\msum{\imu}\wmu\mueffmu^2\var{\Imu}\right)^\frac{1}{2}}%
{\msum{\imu}\wmu\,\mueffmu\tmean{\Imu}}.
\label{e:statisticsxph}
\end{eqnarray} 
where \lp\ is the linear patch size, \sig{\fobs}\ the standard
deviation of the observable flux, and \sig{\Imu} the standard
deviation of the intensity at inclination~\mueffmu. The first equality
in Eq.~\eref{e:statisticsxph} expresses a close relationship between
the level of fluctuations in the photocentric position and the
brightness. This relation carries also over to the temporal power
spectrum of~$\xph(t)$ which is up to a factor $\Rstar^2/6$ identical
to the power spectrum of the brightness fluctuations. Remarkably, the
last equality in Eq.~\eref{e:statisticsxph} expresses that \sig{\xph}\
is \textit{not\/} dependent on stellar radius since only quantities
intrinsic to the simulation enter. The first equality might suggest
the contrary. However, $\sig{\fobs}/\tmean{\fobs}$ scales inversely
proportional to the radius so that the radius factor cancels.  The
similarity in the power spectra has the consequence that the
characteristic time scale for the fluctuations in photocentric
displacement and brightness is the same, namely the evolutionary time
scale of individual convective cells. The similarity in the power
spectra does not imply that the fluctuations are correlated; it can be
shown that photocentric and brightness fluctuations are perfectly
uncorrelated.

\section{Comparison with VIRGO observations}

\begin{figure}[t!]
\begin{center}
\includegraphics[width=\hsize]{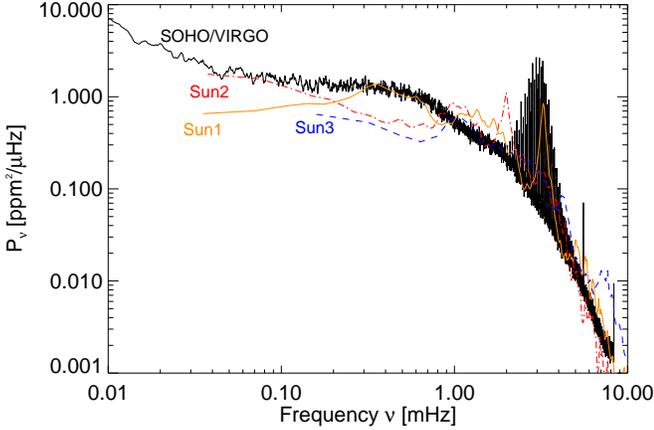}
\caption{Comparison of power spectra of disk-integrated photometric
fluctuations between three solar simulations and observational solar
data from SOHO/VIRGO. Note the steep decline in power in
the range of the p-mode frequencies.\label{f:virgocomp}}
\end{center}
\end{figure}

Figure~\ref{f:virgocomp} shows a comparison between the
disk-integrated, photometric fluctuations derived from our three solar
models, and observational data from the {\sc Virgo} instrument on board
the {\sc Soho} satellite. Our focus is the high frequency region of
the solar signal in which granular contributions dominate. The {\sc
Virgo} power spectrum has been calculated from (level 2) time series
data provided\footnote{
http://www.pmodwrc.ch/pmod.php?topic=tsi/virgo/\\proj\_space\_virgo\#VIRGO\_Radiometry}
by the {\sc Virgo} team. Actually chosen was one year of data
(1996.5-1997.5) close to solar minimum activity to minimise the
possible contribution of activity related variability. Moreover, the
plotted power spectrum is based on data of the green channel of the
{\sc Virgo} three-channel sun-photometer (SPM, see
\cite{Froehlich+al97}). It has been converted to white light
fluctuations by matching (by shifting in power) a corresponding power
spectrum based on {\sc Virgo} PMO6V-A absolute radiometer data in the
frequency range 0.3 to 2.0\pun{mHz}. Instead of using the PMO6V-A
power spectrum directly this rather involved procedure was necessary
since PMO6V-A and SPM power spectra deviate substantially in the high
frequency region. The authors found little information about this
mismatch in the literature. However, it appears to be consensus that
the SPM spectrum reflects the actual solar behaviour
(\cite{Froehlich+al97}, \cite{Andersen+al98}), in particular showing the
steep (roughly as $\nu^{-4}$) decline at the highest frequencies.

We find a reasonable agreement of the continuous background signal
between model predictions and observations between 0.04 and
8.0\pun{mHz}. In particular, the background in the p-mode frequency
range is matched quite well. The differences among the models can be
taken as an estimate of the involved numerical and statistical
theoretical uncertainties. We emphasise that the also visible p-mode
peaks of the models are \textit{not\/} expected to match directly the
observations since the resonance and excitation properties differ
widely between the box models and the Sun.

\section{Trends in the photometric variability}
\label{s:trendphoto}

\begin{figure}[t!]
\begin{center}
\includegraphics[width=0.7\hsize,angle=90]{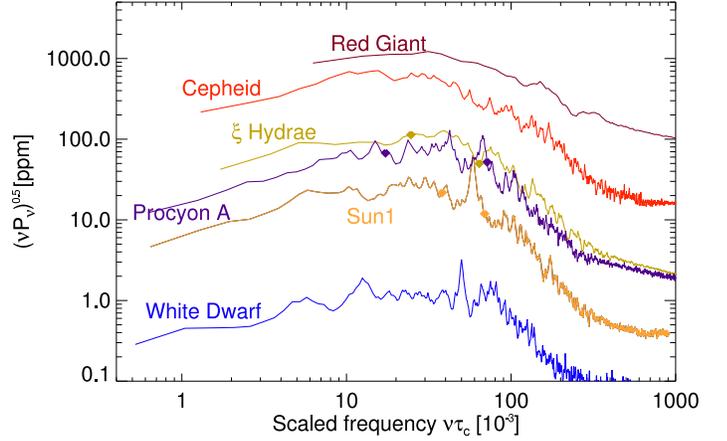}
\caption{Spectral power density of the brightness fluctuations as a
function of scaled frequency $\nu\cdot\tau_\mathrm{c}$. 
Diamonds delimit frequency ranges in which p-modes have been detected
in radial velocity (Procyon data were taken from Brown et al. 1991,
$\xi$~Hydrae data from Frandsen et al. 2002); observed p-mode frequencies
are located in regions with large background power in the models.
Note the systematic increase in power with
decreasing gravity. The behaviour at scaled frequencies greater than
300 is dominated by the numerics and should be
discarded.\label{f:sigftrend}}
\end{center}
\end{figure}

Figure~\ref{f:sigftrend} shows a comparison of the brightness
fluctuations for a sequence of models spanning a range between DA
white dwarfs with convective outer envelopes and red giants. The power
spectral density~$P_\nu$ is plotted as \mbox{$\nu P_\nu$} which makes
it independent of the unit in which the frequency is measured. This
facilitates the intercomparison among the models where we scaled the
frequency with the sound crossing time over a pressure scale height at
the surface (\tauc). As evident from Fig.~\ref{f:sigftrend} all
spectra show a similar shape, and are essentially located in the same
scaled frequency range. We find a systematic increase of the
photometric variability towards giants. Note, that in fact the square
root of the power is plotted in Fig.~\ref{f:sigftrend}. I.e. our
models predict an increase in the \textit{amplitude\/} by a factor of
$\approx 1000$ between the white dwarf and red giant model.

Our result is in marked contrast to the modelling of
\cite*{Trampedach+al98} who did not find an increase in the granular
photometric signal comparing the Sun and Procyon~A. However, they
pointed out that their time series might have been not long enough to
provide sufficient statistics and coverage of lower frequencies. In
view of the ongoing debate about the photometric non-detection of
p-modes in Procyon by the Canadian {\sc Most} satellite
(\cite{Matthews+al04}, \cite{CDK04}) we stress that we obtain a factor
of $\approx 3$ in amplitude between the background signal in the Sun
and Procyon. This refers to a comparison in a representation like
Fig.~\ref{f:sigftrend} where differences in power can be essentially
described by uniform vertical shift. In a representation in absolute
frequencies the power ratio would be frequency-dependent.

\begin{figure}[t!]
\begin{center}
\includegraphics[width=0.65\hsize,angle=90]{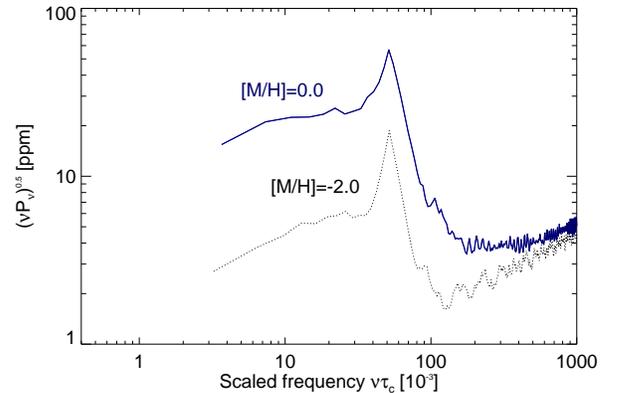}
\caption{Spectral power density of the brightness fluctuations for two
  models around $\Teff=5000\pun{K}$ of solar (filled) and $1/100$
  solar metallicity (dotted). To facilitate comparison the x-axis
  includes the same range of frequencies as the one in
  Fig.~\ref{f:sigftrend}.\label{f:metals}}
\end{center}
\end{figure}

By comparing models of similar effective temperature and surface
gravity, but different metallicity we find a decrease of the
brightness fluctuations with decreasing metallicity, see
Fig.~\ref{f:metals}. This is in line with expectation that the higher
densities encountered at optical depth unity in metal poor models (due
to lower overall opacity) leads to smaller convective fluctuations and
consequently smaller brightness fluctuations.

\section{Trends in the photocentric displacement}

\begin{figure}[t!]
\begin{center}
\includegraphics[width=0.6\hsize,angle=90]{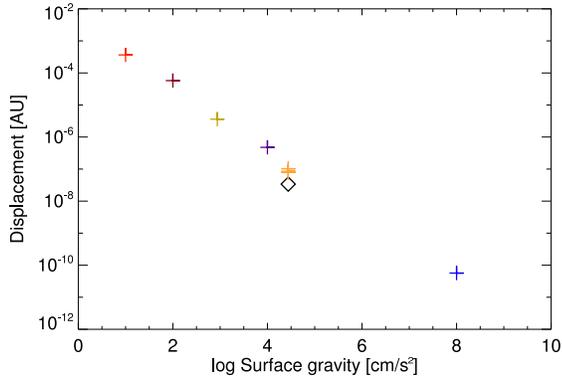}
\caption{Photocentric displacement as a function of
surface gravity. The diamond indicates the metal poor model.\label{f:dis1}}
\end{center}
\end{figure}

We find an almost perfectly linear relationship between the standard
deviation of the photometric displacement and the surface gravity of
the models (see Fig.~\ref{f:dis1}). One has to keep in mind, however,
that there will be some scatter around this line when considering
atmospheres of markedly different effective temperature or chemical
composition from the ones considered here. The diamond depicting the
metal poor model already indicates this.

\begin{figure}[t!]
\begin{center}
\includegraphics[width=0.6\hsize,angle=90]{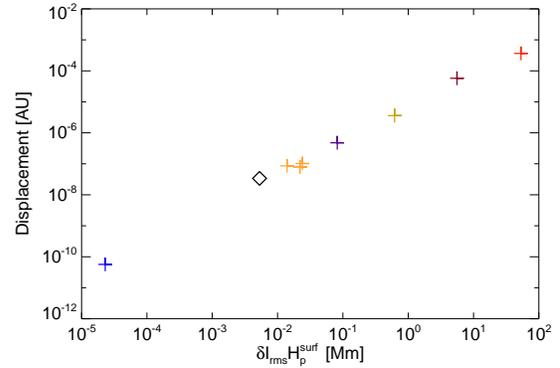}
\caption{The displacement as function of relative
spatial intensity contrast times surface pressure scale height.\label{f:dis2}}
\end{center}
\end{figure}

Instead of the basic atmospheric parameters (\Teff, \logg, chemical
composition), two parameters more closely related to the convection
pattern are perhaps physically better suited to describe the
functional behaviour of the photocentric displacement --- namely the
relative spatial intensity contrast~\dirms\ and the typical size of a
convective cell.  The size of granular cells is related to the
pressure scale height at optical depth unity~\Hpsurf
(\cite{fre01}). Figure~\ref{f:dis2} depicts the outcome of a test of a
relation between photocentric displacement and the product
$\dirms\Hpsurf$: indeed, all models including the metal poor one now
follow the same trend. This also indicates that the proportionality to
$g^{-1}$ seen in Fig.~\ref{f:dis1} is primarily reflecting the
systematic increase of granular cell size with decreasing
gravitational acceleration.

For our most extreme giant model with $\logg=1$ we find a standard
deviation of the photocentric position of $3\cdot 10^{-4}\pun{AU}$ or
$0.3\,\mathrm{mas}/D\left[\mathrm{pc}\right]$ where $D$ denotes the
distance. On a precision level expected to be reached by the {\sc
GAIA} astrometry such a variability does only affect the achievable
precision for close-by giants. The situation for supergiants remains to
be investigated; a linear extrapolation of the curves shown in
Figs.~\ref{f:dis1} and~\ref{f:dis2} towards lower surface gravity is
unlikely to give a reliable estimate of their photocentric variability
since sphericity effects render convection in supergiants markedly
different (\cite{fre02}) from convection in Cartesian geometry as
presented here.

\section{Concluding remarks}

We found systematic changes of the convection-induced fluctuations in
brightness and photocentric displacement with stellar parameters. The
result is a nice example for an application of hydrodynamical model
atmospheres (in this respect see Ludwig \&\ Kucinskas, this volume).
We would like to emphasise again that we have been considering the
contribution of granulation to the stellar variability only. At
highest temporal frequencies it is the dominant contributor to the
variability. However, at lower frequencies variability related to
magnetic activity, spottedness, and rotation dominates, and the level
of the fluctuations is likely to be significantly larger than the
convective contribution. 

\begin{acknowledgements}
The authors would like to thank Matthias Steffen for making available
one of his solar models (Sun3). HGL would like to thank for the lively
as well as insightful discussions with Hans Bruntt and Hans Kjeldsen
on solar and stellar high-precision photometry during a visit to the
astronomy department of the University of Aarhus. This work benefitted
from financial support by the Swedish Research Council and the Royal
Physiographic Society in Lund.
\end{acknowledgements}

\end{document}